# Global Patterns and Drivers of Terrestrial Surface Albedo Dynamics from 2001 to 2025

Xiaoting Huang[1], Zhu Liu[1]

[1] Department of Earth System Science, Ministry of Education Key Laboratory for Earth System Modeling, Institute for Global Change Studies, Tsinghua University, Beijing, 100084, China.

*Correspondence to*: Zhu Liu (zhuliu@tsinghua.edu.cn)

**Broader context**

Land surface albedo shapes how much solar energy the Earth reflects and absorbs, linking land conditions to climate, ecosystems, and water availability. However, a globally integrated understanding of terrestrial albedo change and its environmental drivers remains limited. Using satellite observations and environmental datasets, this study shows that terrestrial albedo change is strongly region-dependent. From 2001 to 2025, significant long-term brightening and darkening coexisted across non-Antarctic land, indicating that recent albedo change was governed by regionally differentiated surface processes rather than a single global tendency. The study further shows that snow, vegetation, and soil water dominate albedo variability in different parts of the world. These patterns matter because changes in land reflectivity can reveal broader shifts in climate, ecosystems, and surface hydrology. Over the longer term, this work supports the development of albedo-based monitoring for environmental change and climate risk assessment. It also highlights the value of collaboration across remote sensing, climate science, ecology, and hydrology to interpret land-surface responses to ongoing warming. By improving the detection of region-specific land-surface change, this research can support climate assessment, ecosystem monitoring, and more informed adaptation, land management, and environmental decision-making.

**Abstract.**

Terrestrial surface albedo is a fundamental determinant of the global radiative balance, yet a spatially integrated understanding of its long-term trajectories and multi-sphere environmental drivers remains lacking. Utilizing MODIS observations (2001–2025), this study integrates Theil–Sen trend analysis, standardized anomaly diagnostics, and a pixel-wise multiple linear regression framework to quantify global albedo dynamics and attribute them to biophysical drivers. Our results reveal that global albedo change is characterized by profound spatial heterogeneity rather than a uniform trend, with 23.3% of the land surface exhibiting significant brightening, while 18.9% showed significant darkening. Multi-driver attribution demonstrates a clear zonal organization of controls: snow cover dominates albedo variability in northern high latitudes (46.9% of classified area), while NDVI (41.0%) and soil moisture (12.1%) govern tropical vegetated regions and moisture-limited transitional zones, respectively. These findings underscore that while terrestrial albedo is a robust indicator of Earth system change, its interpretation depends strongly on regional cryospheric, biophysical, and hydrological contexts.

## 1 Introduction

Terrestrial surface albedo is a fundamental land-surface property that governs how incoming shortwave radiation is partitioned between reflection and absorption[1,2]. Through its direct control on surface radiative exchange, albedo influences surface temperature, boundary-layer development, land–atmosphere coupling, and global climate feedbacks[3,4,5,6]. This property is therefore not merely an optical parameter, but an integrative expression of the land surface biophysical state. Changes in terrestrial albedo can emerge from shifts in vegetation structure and cover, snow and ice conditions, soil exposure, and surface moisture status, linking local land-surface change to broader climate-relevant energy redistribution. This central role of land surface albedo in the surface energy balance and climate feedbacks has been extensively emphasized in both the remote-sensing and climate science literature[1,7]. Therefore, clarifying the complex dynamics of albedo across different spatial and temporal scales under multi-process coupling is a central issue in understanding land surface processes and climate feedbacks.

The dynamics of terrestrial surface albedo are inherently heterogeneous because the processes controlling albedo vary strongly across biomes, climates, and seasons[8]. In densely vegetated ecosystems, albedo is closely related to canopy structure, leaf area, background soil masking, and phenological change[9]. In drylands, albedo often responds to the delicate balance between sparse vegetation cover, bare-soil exposure, and soil wetness, and may therefore shift substantially during drought development or land degradation[10,11]. In high-latitude regions, albedo is strongly modulated by snow cover, snow persistence, and vegetation snow-masking effects, making the snow–albedo feedback one of the most potent biophysical mechanisms in the climate system[12]. The climatic importance of forest-albedo effects, snow-albedo feedbacks, and vegetation-mediated radiative changes has been demonstrated across a wide range of studies, from boreal afforestation impacts to global greening analyses [7,13].

These spatially variable controls also mean that the same sign of albedo change can imply fundamentally different underlying surface processes. A decrease in albedo may indicate canopy densification and enhanced vegetation cover in humid or temperate regions, while in snow-dominated regions, it may reflect stronger snow masking by expanding shrubs and forests, or reduced snow persistence under regional warming[14,15]. Conversely, albedo increases in arid and semi-arid environments may be associated with vegetation decline, enhanced bare-soil exposure, or severe deficits in soil moisture. Empirical and observational studies have consistently shown that bare-soil albedo is highly sensitive to soil water content[16], while broad-scale albedo variability can covary with snow cover, vegetation dynamics, and hydroclimatic factors. This biome-dependent interpretability makes terrestrial albedo scientifically informative, but it also necessitates that global-scale diagnosis be framed carefully to avoid physical misinterpretation.

Satellite observations now provide an unprecedented capability for monitoring terrestrial albedo over long periods and across the globe. In particular, the MODIS BRDF/albedo framework established a consistent, physically based algorithm for the global retrieval of land-surface albedo[17,18], and subsequent validation studies have confirmed that these products are sufficiently accurate for rigorous scientific analysis under varied conditions[19]. At the same time, recent reviews have emphasized that despite major progress in retrieval algorithms, spatial gap-filling, and cross-sensor harmonization, a globally integrated understanding of land-surface albedo dynamics across different land types and climate regimes remains incomplete[20]. Many previous studies have focused on regional changes, specific land-cover transitions, or individual process domains, whereas fewer studies have holistically combined long-term albedo trajectories, recent extreme anomalies, and multivariate environmental drivers within a single observational framework.

Placing recent anomalies in the context of long-term change is especially critical for a physically meaningful interpretation of present-day land-surface dynamics. Long-term trend analysis identifies whether land surfaces have undergone persistent brightening or darkening, thereby establishing the background trajectory of terrestrial radiative change. Anomaly analysis, by contrast, isolates short-term departures from climatological expectations and is highly sensitive to recent disturbances, climate variability, and extreme hydroclimatic events. This trend–anomaly combination has proven highly effective in recent large-scale monitoring studies, interpreting the latest focal year against a multidecadal baseline [21,22]. For terrestrial surface albedo, such a framework can definitively reveal whether recent conditions reinforce, offset, or completely diverge from longer-term land-surface trajectories.

Beyond documenting spatiotemporal patterns, explicitly attributing the drivers of albedo dynamics is essential for robust interpretation. Because albedo integrates the complex surface expressions of vegetation, hydroclimate, and cryospheric processes, both long-term trajectories and recent anomalies are driven by multiple interacting controls rather than a single isolated factor. Here, we use satellite-derived terrestrial shortwave albedo observations to quantify long-term spatiotemporal trends (2001–2025) using the robust Theil–Sen estimator and to explicitly characterize standardized anomalies and historical rankings with a specific focus on 2025. Specifically, we evaluate the underlying environmental drivers utilizing a multiple linear regression framework. To characterize the environmental factors associated with terrestrial surface albedo variability, we incorporate three variables representing cryospheric (snow cover), biophysical (Normalized Difference Vegetation Index, NDVI), and hydrological (soil moisture) processes. By employing standardized regression coefficients and rigorous multicollinearity diagnostics, we aim to systematically quantify the relative contribution and spatial dominance of each environmental driver on albedo perturbations. Ultimately, by integrating long-term trend trajectories, extreme anomaly diagnostics, and multivariate driver attribution, this study establishes terrestrial surface albedo as a physically interpretable indicator for global climate and land-surface monitoring.

## 2 Data and Method

### 2.1 Satellite observations of terrestrial surface albedo

This study used the MODIS MCD43C3 (Collection 6.1) daily surface albedo product at 0.05° spatial resolution[23]. Given its direct relevance to the surface energy budget, the shortwave (0.3–5.0 μm) White-Sky Albedo (WSA) was selected as the main variable for analysis. To ensure the consistency and reliability of the pixel-level time series, a set of Quality Assurance and Quality Control (QA/QC) procedures was applied before analysis. First, only pixels with good retrieval quality (quality flag ≤ 2) were retained to reduce noise associated with lower-quality inversions. Second, pixels were required to satisfy the criterion of Percent_Inputs ≥ 50, ensuring that the albedo retrieval was supported by a sufficient number of valid clear-sky observations within the synthesis period. After daily pixel-level filtering, the retained observations were aggregated into monthly composites using the pixel-wise mean. This temporal aggregation further reduced short-term variability and provided a more stable basis for subsequent trend and anomaly analyses. For the regression analysis, monthly data were upscaled to a 0.25° spatial grid via area-weighted averaging to match the resolution of the meteorological driving datasets.

### 2.2 Environmental Driving Factors

To disentangle the diverse mechanisms governing surface albedo dynamics, three independent datasets representing distinct cryospheric, hydrological, and biophysical controls were integrated into this study. First, to capture the dominant cryospheric forcing, snow cover fraction data were obtained from the MOD10CM product[24]. Second, volumetric soil moisture data for the surface layer from ERA5-Land were utilized to quantify the hydrological status of the land surface[26], which physically alters the optical properties of bare soils through soil darkening effects. Third, the Normalized Difference Vegetation Index (NDVI) derived from the MODIS MCD43C4 product[27] was employed as a proxy for vegetation canopy dynamics and structural phenology. All auxiliary datasets were temporally aggregated to monthly means and spatially resampled to the consistent 0.25° grid utilizing a conservative area-weighted averaging approach prior to statistical analyses.

### 2.3 Long-term Trend Analysis

To quantify the long-term temporal evolution of surface albedo, annual mean albedo was first calculated for each pixel. Long-term trends were then estimated using the Theil–Sen (TS) median slope estimator, and their statistical significance was assessed with the Mann–Kendall (MK) non-parametric test[25]. This approach is widely used for remote sensing time series because it is less sensitive to noise and extreme values than ordinary least squares regression and does not require the assumption of normality.

For each pixel, the magnitude of the long-term trend ($\beta$) was estimated using the Theil–Sen median slope:

$$\beta = \text{Median}\left(\frac{x_j - x_i}{j - i}\right), \forall 1 \leq i < j \leq n \tag{1}$$

where $x_i$ and $x_j$ represent the albedo observations in years $i$ and $j$, respectively, and $n$ is the length of the time series. A positive $\beta$ indicates an increasing trend in albedo over time, whereas a negative $\beta$ indicates a decreasing trend.

The statistical significance of the TS-derived trend was then evaluated using the Mann–Kendall test, which detects monotonic changes in a time series without assuming a specific data distribution. The MK test statistic ($S$) is defined as:

$$S = \sum_{i=1}^{n-1} \sum_{j=i+1}^{n} sgn(x_j - x_i) \tag{2}$$

where sgn is the sign function:

$$sgn(\theta) = \begin{cases} 1, & \theta > 0 \\ 0, & \theta = 0 \\ -1, & \theta < 0 \end{cases} \tag{3}$$

For time series with $n \geq 10$, the statistic $S$ approximately follows a normal distribution. The standardized test statistic ($Z$) is therefore calculated as:

$$Z = \begin{cases} \frac{S-1}{\sqrt{Var(S)}}, & S > 0 \\ 0, & S = 0 \\ \frac{S+1}{\sqrt{Var(S)}}, & S < 0 \end{cases} \tag{4}$$

Based on the standardized $Z$ value, pixel-level trends were classified as significant or non-significant. In this study, a significance level of $p < 0.05$ was adopted, corresponding to $|Z| > 1.96$.

## 2.4 Spatiotemporal Anomaly Extraction

To quantify the magnitude and spatial distribution of surface albedo anomalies, a month-specific anomaly framework was adopted. This approach was necessary because surface albedo exhibits a pronounced seasonal cycle, and direct comparison against a single multi-year mean would mix true anomalies with normal seasonal variability. By evaluating each calendar month relative to its own historical baseline, anomaly estimates more accurately capture departures from the expected seasonal state.

For each pixel, the long-term mean ($\mu_m$) and standard deviation ($\sigma_m$) were calculated separately for each calendar month $m$ using the snow-minimized monthly albedo record over the reference period. The standardized anomaly for month $m$ in year $t$ was then defined as:

$$Z_{m,t} = \frac{\alpha_{m,t} - \mu_m}{\sigma_m} \tag{5}$$

where $\alpha_{m,t}$ represents the observed monthly mean albedo for month $m$ in year $t$. Positive values indicate that surface albedo was higher than the historical average for that month, whereas negative values indicate lower-than-normal conditions. This standardization expresses anomaly magnitude relative to the historical variability of each pixel and therefore improves comparability across regions with different background albedo regimes. To summarize anomaly conditions at broader spatial scales, pixel-level monthly anomalies were further aggregated to regional and global means using area-weighted averaging based on the cosine of latitude. This weighting accounts for the latitudinal decrease in pixel area under the geographic coordinate system and ensures that large-scale averages are not biased toward high-latitude grid cells.

To further place anomalies in a long-term context, a temporal ranking analysis was applied to the aggregated regional and global anomaly series. Monthly and annual anomaly values were ranked within the full observational record to determine the relative position of individual years and months. This procedure provides a direct measure of whether anomaly conditions during a given period were among the highest or lowest in the historical record and allows large-scale anomaly events to be interpreted within the broader temporal context of surface albedo variability.

**2.5 Regression analysis of monthly surface albedo and environmental drivers**

To quantify the relative associations of terrestrial surface albedo with major environmental drivers, we performed a pixel-wise standardized multiple linear regression using raw monthly surface albedo, snow cover, soil water, and NDVI at 0.25° monthly resolution, as monthly albedo variability is shaped by interacting cryospheric, hydrological, and biophysical processes. For each land pixel, the model was expressed as:

$$y_t = \beta_0 + \beta_1 SC_t + \beta_2 SW_t + \beta_3 NDVI_t + \varepsilon_t \tag{6}$$

where $y_t$ denotes monthly surface albedo, $SC_t$, $SW_t$, and $NDVI_t$ denote monthly snow cover, soil water, and NDVI, respectively, and $\varepsilon_t$ is the residual term. To enable direct comparison of coefficient magnitudes within each pixel, the response and predictors were standardized along the time dimension before fitting. Regression was conducted independently for each pixel using ordinary least squares. Only predictors with sufficient temporal variation and data availability were retained in the regression. Statistical significance was evaluated at $p < 0.05$. For each pixel, the dominant associated driver was identified as the significant predictor with the largest absolute standardized coefficient, while pixels without any significant predictor were not assigned to a dominant class.

**3 Result**

**3.1 Climatological patterns and long-term trends in terrestrial surface albedo**

The climatological distribution of terrestrial surface albedo showed a clear and physically coherent spatial structure (Fig. 1). Annual mean albedo was generally low over humid and densely vegetated regions, particularly across the humid tropical forest belt and southeastern Asia, and relatively high over arid lands and snow-affected high-latitude regions, especially the Sahara, Arabian Peninsula and northern snow-affected land areas. This contrast produced a pronounced zonal pattern[23], with a broad low-albedo belt in the humid tropics and progressively higher values toward the subtropics and northern high latitudes. Seasonal maps further showed that this pattern was not static. In boreal winter, albedo increased markedly across boreal Eurasia and North America, consistent with enhanced snow exposure, whereas it declined substantially in summer as snow cover retreated and vegetation became more exposed[28]. By contrast, tropical regions exhibited relatively weak seasonal variation, while transition zones showed stronger seasonal contrasts, indicating tighter seasonal coupling of albedo to changes in vegetation cover, soil exposure, and surface wetness.

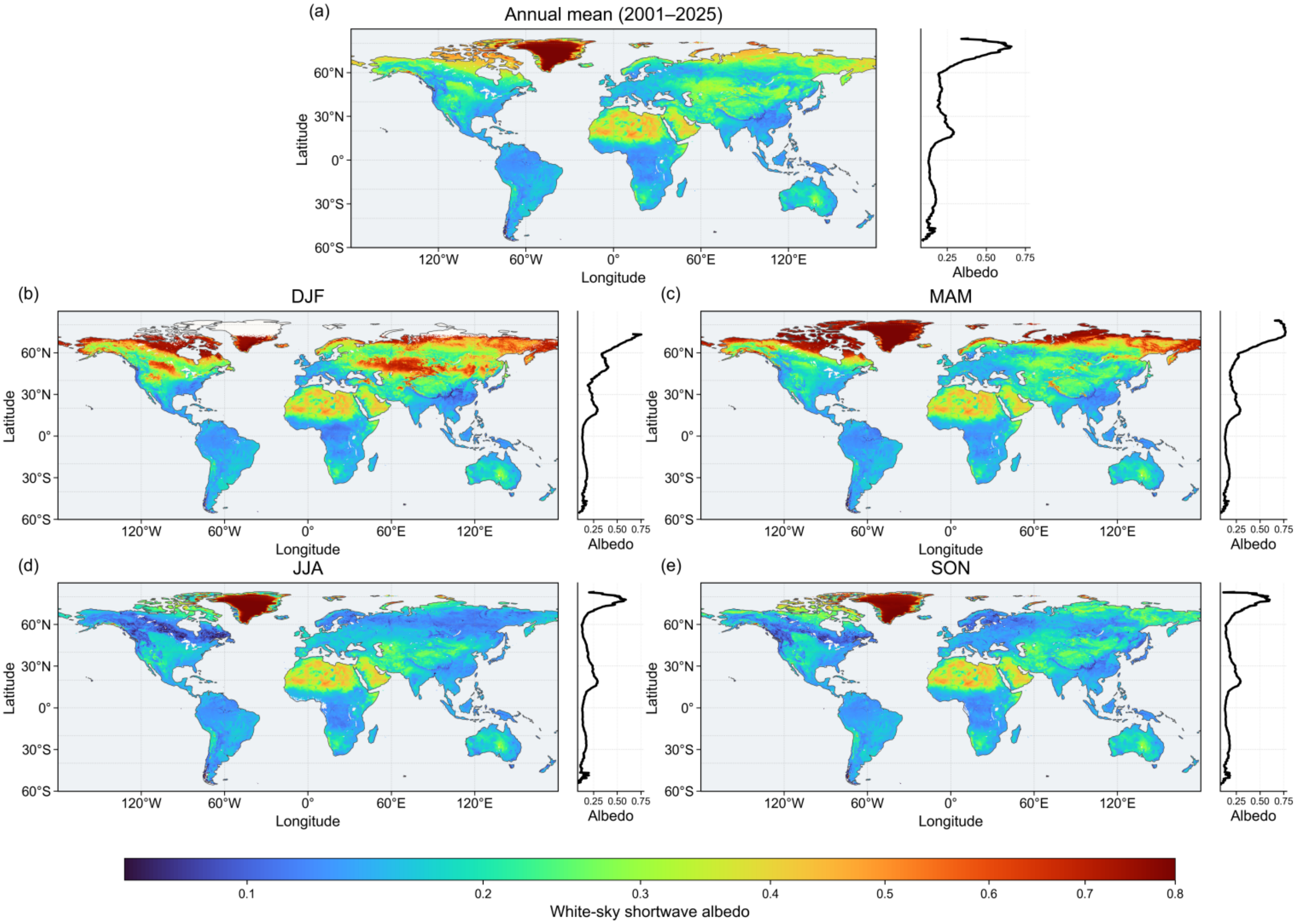

**Figure 1. Climatological spatial patterns and latitudinal profiles of global terrestrial surface albedo (2001–2025). (a)** Spatial distribution of the annual mean white-sky shortwave albedo and its corresponding zonal mean profile (right panel). **(b–e)** Seasonal climatological mean spatial patterns and their respective zonal average profiles for **(b)** December–February (DJF), **(c)** March–May (MAM), **(d)** June–August (JJA), and **(e)** September–November (SON). In each sub-panel, the right-side line chart shows the latitudinal variation of the zonally averaged albedo.

Against this climatological background, long-term trends during 2001–2025 were strongly spatially heterogeneous rather than globally uniform (Fig. 2a). Significant decreases in surface albedo were concentrated mainly in the Indian subcontinent, central Eurasia, and parts of temperate Europe and North America, whereas significant increases were more common in the Arabian

Peninsula and surrounding drylands, parts of tropical South America, and patches of North America and East Asia. These opposing regional signals indicate that recent albedo change was dominated by geographically differentiated surface processes rather than by a single coherent global tendency. Overall, 23.3% of non-Antarctic land pixels showed significant positive trends, 18.9% showed significant negative trends, and the remainder (57.8%) showed no significant long-term change (Fig. 2b). Rather than being evenly distributed, these trend pixels were concentrated in specific climatic zones, highlighting the spatial selectivity of long-term terrestrial albedo change.

The distribution of significant trends showed a clear latitudinal asymmetry. Positive trends were concentrated mainly in low-latitude regions, whereas negative trends were distributed primarily across low- and mid-latitude regions. This pattern indicates that long-term albedo change was structured more strongly by broad climatic zones than by a spatially uniform global tendency. The plant functional type composition showed distinct distributional features among significant trend classes (Fig. 2c). In low-latitude regions, significant positive trends were most strongly represented in grassland systems (39.8%), whereas in mid-latitude regions they were most strongly represented in bare-soil systems (41.0%). For significant negative trends, bare soil accounted for the largest share in low latitudes (44.3%), while forest was the most represented class in mid latitudes (34.9%). At the aggregate level, positive-trend pixels were dominated by grassland (33.2%), whereas negative-trend pixels were dominated by bare soil (33.4%). Overall, these results indicate that the dominant surface background associated with significant albedo trends varied with both latitude band and trend sign, while each trend class still retained a mixed PFT composition.

Seasonal decomposition of annual significant pixels showed uneven seasonal contributions to the annual albedo trend pattern. Annual positive-trend pixels were most strongly associated with DJF (37.9%), whereas annual negative-trend pixels were more strongly associated with MAM (36.0%). JJA contributed the smallest share to both positive (17.9%) and negative (15.1%) annual trend areas. Spatially, MAM-dominated negative trends were concentrated mainly in northern high latitudes, while DJF-dominated negative trends were more evident over parts of South America and high-latitude North America. These results indicate that the annual trend pattern arose from seasonally differentiated contributions that also varied geographically.

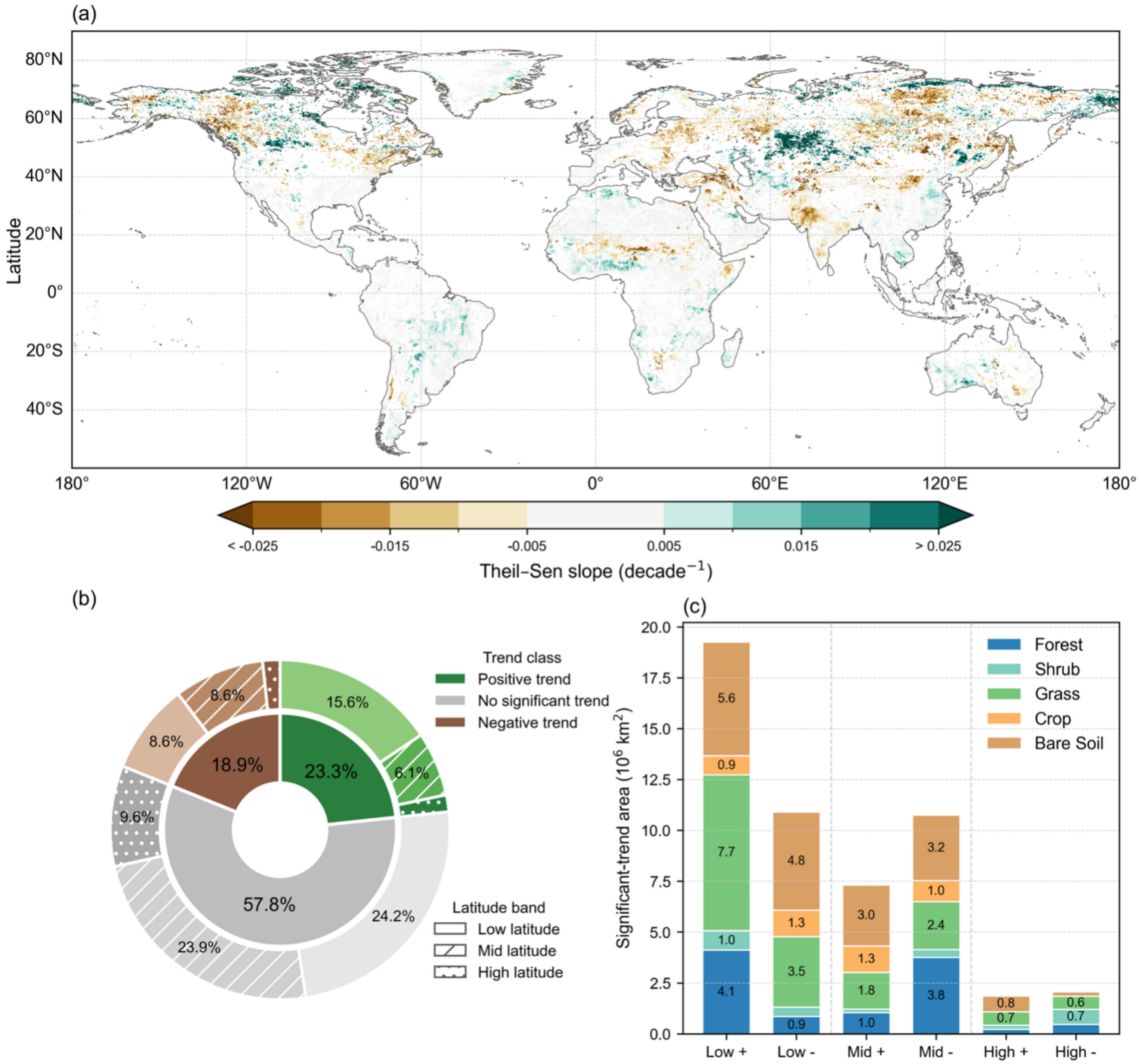

**Figure 2. Long-term trends in global terrestrial surface albedo and their latitudinal and PFT composition. (a)** Spatial distribution of long-term albedo trends across non-Antarctic terrestrial surfaces, estimated using the Theil–Sen median slope (decade$^{-1}$). Only pixels with statistically significant trends based on the Mann–Kendall test ($p < 0.05$) are colored, whereas all other pixels, including those with non-significant trends and those with insufficient valid data for regression, are shown in white. **(b)** Donut chart showing the area-weighted fractions of non-Antarctic land pixels with positive trends, negative trends, and no significant trends, with the outer ring indicating the contribution of low-, mid-, and high-latitude zones. **(c)** Stacked bar

chart showing the absolute area of significant positive and negative trend pixels in low-, mid-, and high-latitude regions, separated by merged plant functional type groups.

### 3.2 Spatial and temporal anomalies in terrestrial surface albedo

Terrestrial surface albedo anomalies exhibited pronounced interannual and spatial variability during 2001–2025 (Fig. 3). The latitude–time distribution showed several coherent anomaly clusters over the Southern Hemisphere subtropics and mid-latitudes, particularly around 20–40°S, including a strong negative anomaly around 2001, another negative phase around 2011–2012, and a pronounced positive anomaly around 2019–2020. In contrast, Northern Hemisphere anomalies were generally weaker and more spatially fragmented. Annual hemispheric anomaly series further showed that the Southern Hemisphere dominated the global interannual signal, with its most negative and most positive annual anomalies occurring in 2001 (−0.0062) and 2019 (0.0052), respectively, whereas Northern Hemisphere anomalies remained much smaller in amplitude, ranging from −0.0016 in 2004 to 0.0016 in 2011. The Southern Hemisphere anomaly clusters were associated with regionally coherent patterns rather than spatially uniform hemispheric shifts. The negative phase around 2011–2012 was expressed most clearly over Australia and parts of southern Africa, while much weaker or mixed signals were evident elsewhere in the Southern Hemisphere. By contrast, the positive phase around 2019–2020 was more spatially consistent, with widespread positive anomalies across Australia, southern Africa, and parts of southern South America. This contrast indicates that the negative and positive Southern Hemisphere anomaly phases differed not only in sign but also in their spatial coherence across major southern land regions.

Annual anomalies in 2025 were spatially heterogeneous rather than globally coherent (Fig. S2). Negative anomalies were concentrated mainly over South Asia and Australia, whereas positive anomalies were more evident across northern extratropical Eurasia and North America[29,30], including parts of Europe–western Eurasia, Central Asia, northern Eurasia, and the North American boreal zone. Elsewhere, including southern Africa, southern South America, and the Amazon, annual anomalies were comparatively weak. Together, these patterns indicate that 2025 was characterized by strong regional contrasts rather than by a coherent global-scale shift.

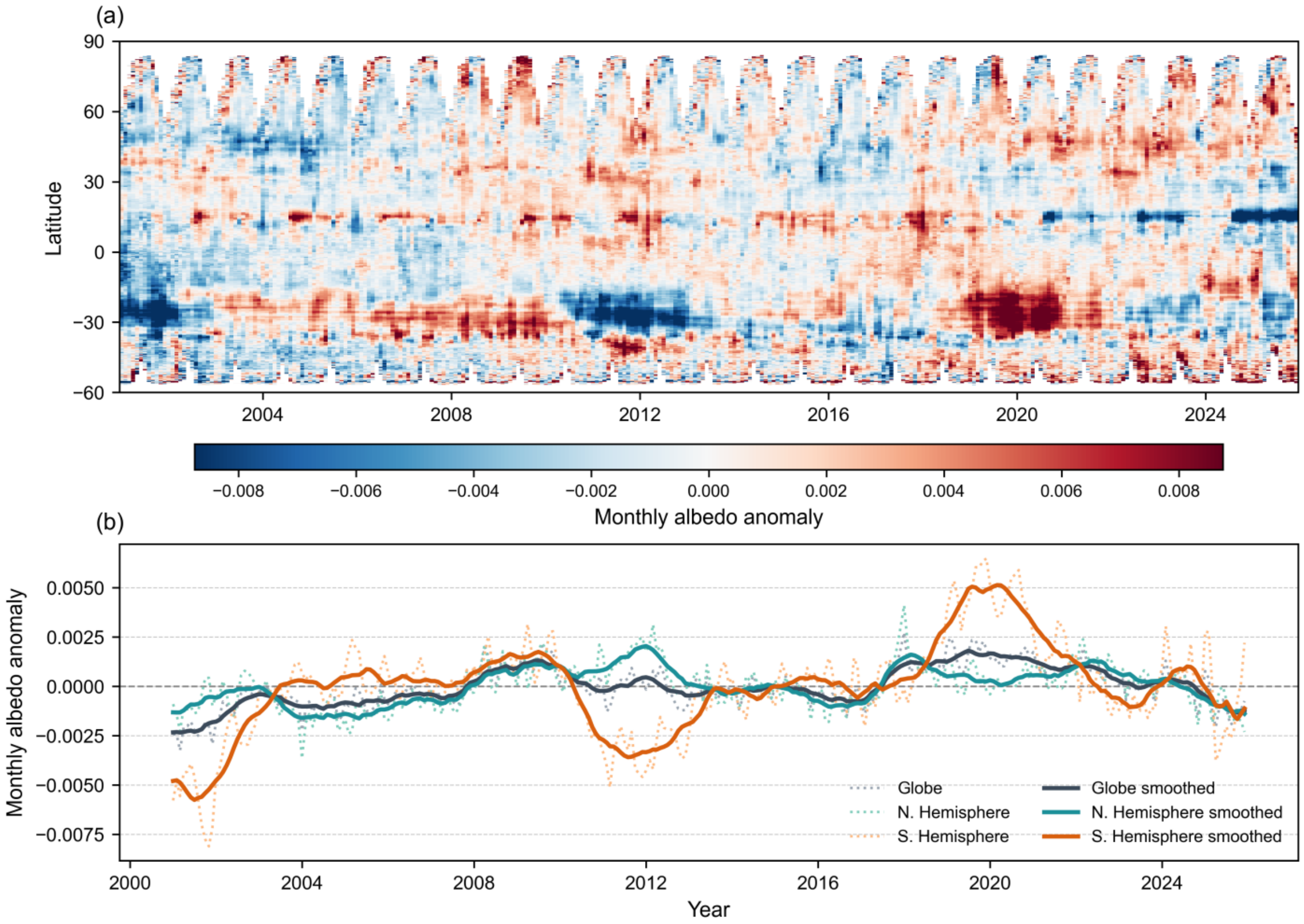

**Figure 3. Spatiotemporal variability of terrestrial surface albedo anomalies during 2001–2025. (a)** Latitude–time distribution of monthly terrestrial surface albedo anomalies relative to the climatological mean. **(b)** Global, Northern Hemisphere, and Southern Hemisphere mean monthly albedo anomalies. Dashed lines indicate the original monthly series, and solid lines indicate the smoothed series.

### 3.3 Relationships between terrestrial surface albedo and environmental drivers

Pixel-wise standardized multiple regression revealed strong regional contrasts in the dominant relationships between monthly terrestrial surface albedo and environmental drivers (Fig. 4a). In the main dominance map, only pixels with $R^2 \geq 0.30$ are shown. Among these pixels, snow was the dominant correlate across most northern mid- to high-latitude land areas, forming extensive and spatially coherent zones over boreal North America and northern Eurasia, and accounting for 46.9% of the classified land area. NDVI dominated much of the lower-latitude vegetated land surface, including large parts of Africa, South America, South and East Asia, and Australia, representing 41.0% of the classified land area. Soil water accounted for the remaining 12.1% and was concentrated mainly in moisture-limited transition zones, including western North America, the Mediterranean–West Asia belt, southern South America, southern Africa, and parts of Australia. These patterns show that the dominant environmental relationships of monthly terrestrial albedo vary strongly across regions.

The area-weighted latitudinal composition of dominant correlates further highlights a clear zonal transition (Fig. 4b). Snow dominated northern high-latitude bands, accounting for 87.6% and 70.0% of classified land area in 45–60°N and 60–75°N, respectively. By contrast, tropical bands were characterized primarily by NDVI dominance, with NDVI representing 77.9% of classified land area in 0–15°N and 79.3% in 15°S–0°. In the Southern Hemisphere mid-latitudes, dominance shifted toward a combination of soil water and NDVI: in 30–45°S, soil water and NDVI accounted for 57.2% and 37.5% of classified land area, whereas snow contributed only 5.3%. However, the fraction of land area meeting the display criterion was much lower in equatorial bands than in the extratropic, reaching only 42.7% in 0–15°N and 23.1% in 15°S–0°, compared with 96.9% in 45–60°N, 93.4% in 60–75°N, and 81.2% in 30–45°S. This contrast indicates that the selected predictors captured monthly albedo variability much more effectively in extratropical than in humid tropical regions.

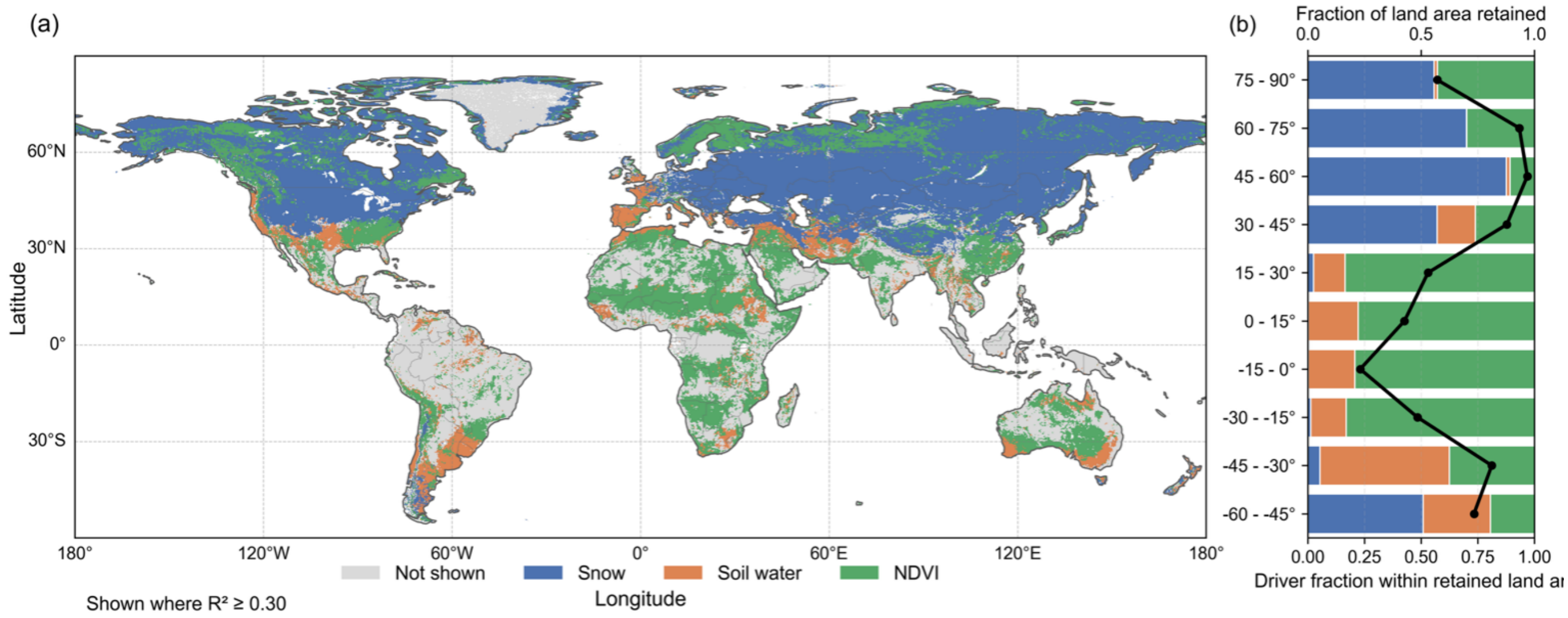

**Figure 4. Spatial dominance and latitudinal composition of the main environmental correlates of monthly terrestrial surface albedo. (a)** Global distribution of the dominant environmental correlate of raw monthly terrestrial surface albedo. **(b)** Area-weighted latitudinal composition of dominant environmental correlates.

The coefficient maps provide further insight into the direction and spatial structure of these relationships (Fig. S2a–d). Snow cover showed predominantly positive standardized coefficients across northern mid- and high-latitude regions, with 90.1% of significant snow coefficients being positive[31], consistent with the strong brightening effect of increasing snow cover on land-surface reflectivity. Soil water coefficients were predominantly negative, with only 18.8% of significant pixels showing positive associations, consistent with soil darkening under wetter surface conditions in many moisture-limited regions[32]. NDVI coefficients were more spatially heterogeneous, with 40.5% of significant pixels showing positive associations[33], indicating that the vegetation–albedo relationship does not follow a spatially uniform sign. Spatial differences in model fit further support this pattern: mean $R^2$ values were 0.75 and 0.76 in 45–60°N and 60–75°N, respectively, but only 0.31 in 0–15°N and 0.20 in 15°S–0°, with an intermediate value of 0.47 in 30–45°S. Together, these results indicate that monthly terrestrial surface albedo reflects distinct cryospheric, hydrological, and biophysical regimes whose relative importance varies systematically across latitude and environmental setting.

## 4. Discussion

Terrestrial surface albedo changes during 2001–2025 were characterized primarily by strong spatial heterogeneity rather than by a globally uniform trend. Brightening and darkening co-occurred across major climatic zones, indicating that albedo change reflects regionally distinct land-surface processes. This highlights that large-scale albedo trends should be interpreted in their environmental context, rather than as a single coherent signal. Combining long-term trends with recent anomalies provides a clearer view of land-surface change. Long-term trends describe the background evolution of surface reflectivity, whereas anomalies capture short-term departures from that background[34]. The 2025 pattern was spatially mixed rather than globally coherent, suggesting that recent variability was dominated by regional hydroclimatic and ecological perturbations instead of a synchronized global shift.

The driver analysis further showed a clear geographical organization of monthly albedo variability. Snow was the dominant factor across northern mid- to high-latitude regions[35], NDVI dominated much of the lower-latitude vegetated land surface, and soil water was most influential in moisture-limited transitional zones. These results indicate that global albedo variability is structured by distinct cryospheric, biophysical, and hydrological regimes. Lower explanatory power in humid tropical regions suggests that albedo variability there is governed by more complex interacting processes than those represented by the selected predictors. Overall, this study shows that terrestrial albedo is a useful indicator of land-surface change, but its interpretation must account for strong regional heterogeneity in both patterns and controls.

## 5. Conclusions

Using satellite observations and ancillary environmental datasets, this study investigated terrestrial surface albedo dynamics during 2001–2025 from the perspectives of long-term trends, recent anomalies, and environmental controls. The results show that terrestrial surface albedo exhibited pronounced spatial heterogeneity rather than a globally uniform pattern. Across non-Antarctic land, 23.3% of pixels showed significant positive trends, 18.9% showed significant negative trends, and 57.8% showed no significant long-term change, indicating that recent albedo change was dominated by regionally differentiated land-surface processes.

Recent anomaly analysis showed that albedo conditions in 2025 were also spatially heterogeneous. Negative anomalies were concentrated mainly over South Asia and Australia, whereas positive anomalies were more evident across northern extratropical Eurasia and North America. These patterns suggest that recent albedo variability was expressed primarily through regional contrasts rather than a coherent global-scale shift.

The relationships between albedo and environmental drivers also showed a clear geographical structure. Among pixels with $R^2 \geq 0.30$, snow, NDVI, and soil water accounted for 46.9%, 41.0%, and 12.1% of the classified land area, respectively. Snow was the dominant control in northern high latitudes, NDVI dominated tropical vegetated regions, and soil water was most important in moisture-limited transitional zones. Together, these results indicate that terrestrial albedo is closely linked to cryospheric, biophysical, and hydrological processes, and that its interpretation requires explicit consideration of regional environmental context.

**Supplement**

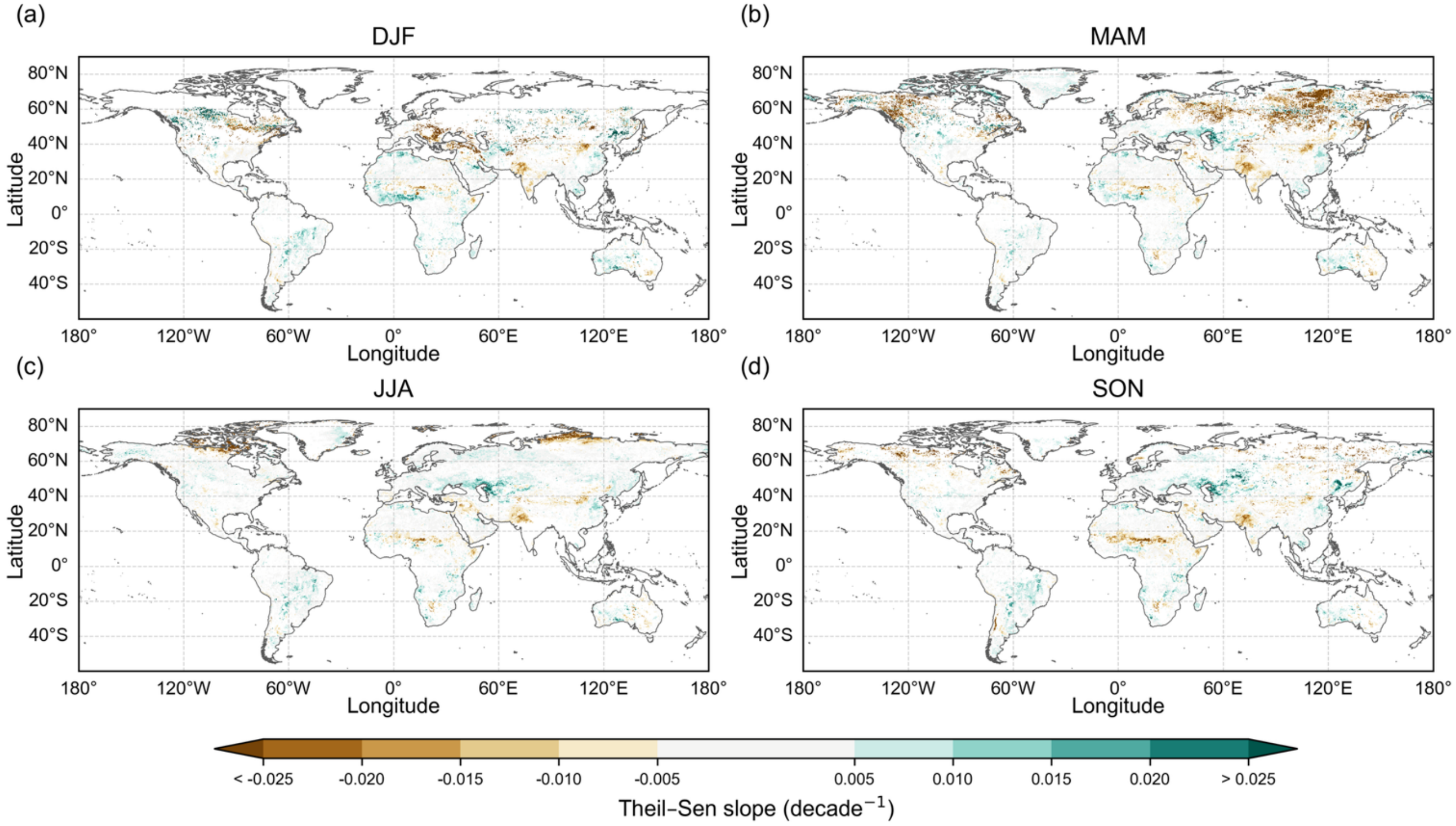

**Figure S1. Seasonal patterns of long-term trends in terrestrial surface albedo. (a–d)** Spatial distribution of pixel-wise albedo trends in DJF, MAM, JJA, and SON, estimated using the Theil–Sen slope (decade$^{-1}$). Only pixels with statistically

significant trends based on the Mann–Kendall test are shown in color. All other pixels, including those with non-significant trends and those with insufficient valid data for regression, are shown in white.

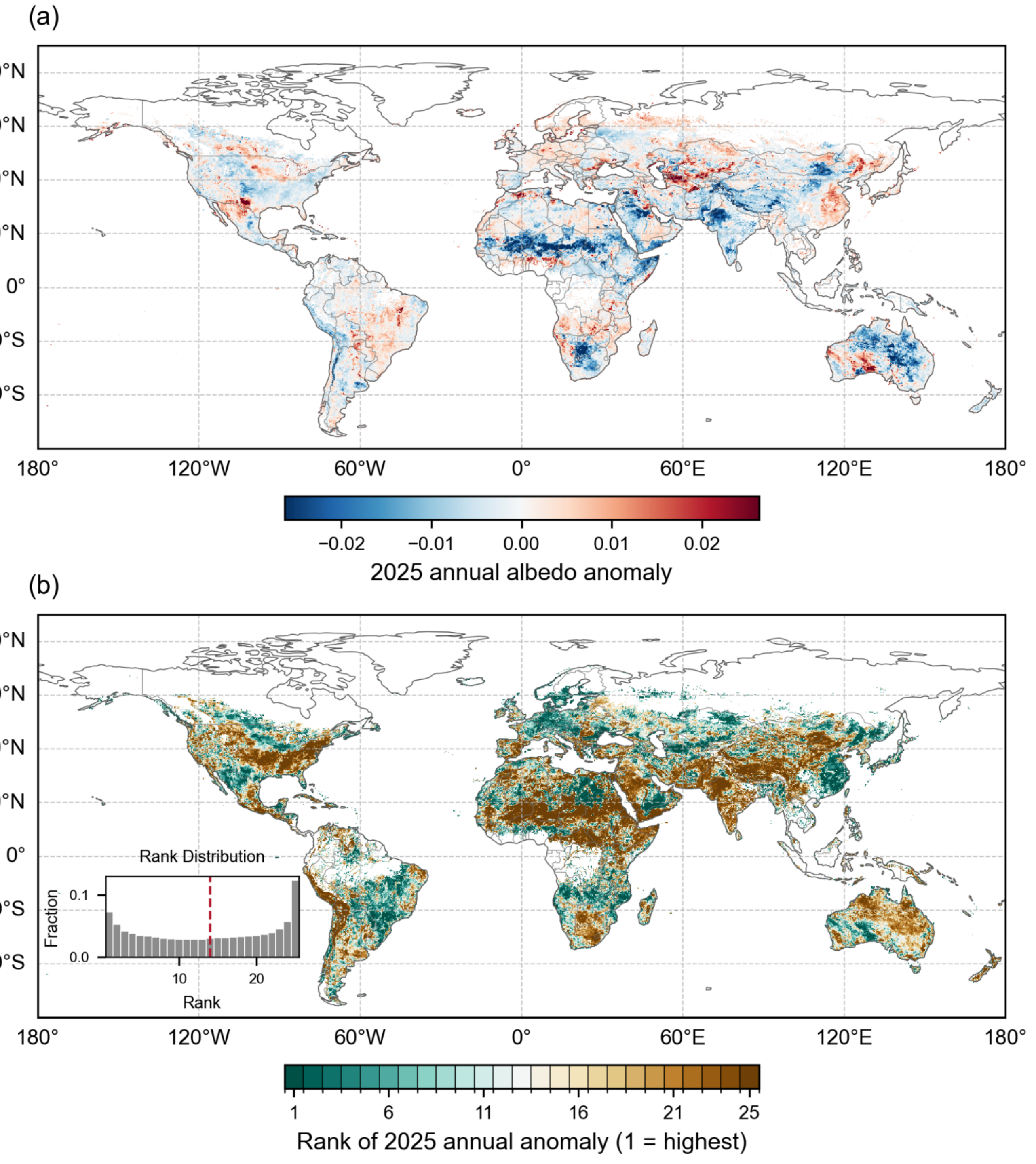

**Figure S2. Spatial distribution and historical ranking of terrestrial surface albedo anomalies in 2025. (a)** Annual terrestrial surface albedo anomalies in 2025 relative to the 2001–2025 climatology. **(b)** Historical rank of the 2025 annual albedo anomaly at each grid cell within the 2001–2025 record. The inset shows the frequency distribution of grid-cell ranks.

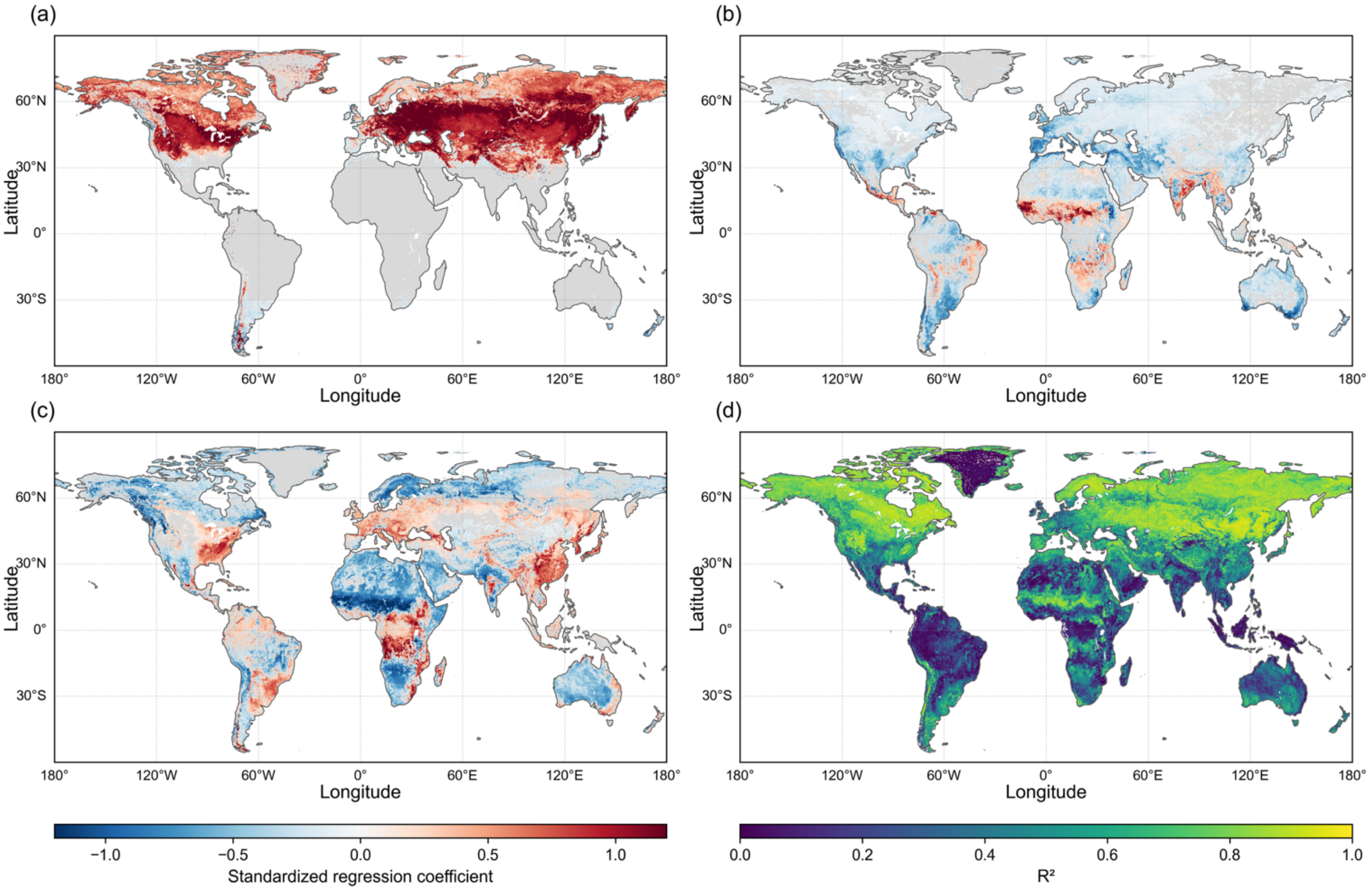

**Figure S3. Spatial distribution of significant standardized regression coefficients and model fit for monthly albedo relationships. (a–c)** present the significant standardized coefficients for snow cover**,** soil water**,** and NDVI, respectively, obtained from pixel-wise raw monthly regressions of terrestrial surface albedo. Only coefficients significant at $p < 0.05$ are shown and non-significant coefficients are shown in light gray. **(d)** shows the corresponding $R^2$.

## Author contributions

X.H. conceived the study, performed the data processing and formal analysis, prepared the figures, and wrote the original manuscript. Z.L. supervised the study, provided conceptual guidance, secured funding, and revised the manuscript.

## Competing interests

The authors declare no competing interests.

## Data Availability Statement

The MODIS MCD43C3 product (Version 6.1) is available from the NASA Land Processes DAAC at https://doi.org/10.5067/MODIS/MCD43C3.061 (last accessed 31 March 2026). The MOD10CM product is available from the NASA NSIDC DAAC at https://nsidc.org/data/mod10cm (last accessed 31 March 2026). The ERA5-Land data are available from the Copernicus Climate Data Store at https://cds.climate.copernicus.eu/ (last accessed 31 March 2026).

## Acknowledgements

The authors thank Dr. Yuhan Yu for her assistance with the introduction, manuscript formatting, language editing, and reference preparation.

## References

1. Bonan, G.B. (2008). Forests and Climate Change: Forcings, Feedbacks, and the Climate Benefits of Forests. Science *320*, 1444–1449. https://doi.org/10.1126/science.1155121.

2. Stephens, G.L., O'Brien, D., Webster, P.J., Pilewski, P., Kato, S., and Li, J. (2015). The albedo of Earth. Reviews of Geophysics *53*, 141–163. https://doi.org/10.1002/2014RG000449.

3. Dickinson, R.E. (1983). Land Surface Processes and Climate—Surface Albedos and Energy Balance. In Advances in Geophysics (Elsevier), pp. 305–353. https://doi.org/10.1016/S0065-2687(08)60176-4.

4. Trenberth, K.E., Fasullo, J.T., and Kiehl, J. (2009). Earth's Global Energy Budget. Bull. Amer. Meteor. Soc. *90*, 311–324. https://doi.org/10.1175/2008BAMS2634.1.

5. Charney, J., Stone, P. H., & Quirk, W. J. (1975). Drought in the Sahara: a biogeophysical feedback mechanism. Science, 187(4175), 434-435.

6. Dickinson, R.E. (1983). Land Surface Processes and Climate—Surface Albedos and Energy Balance. In Advances in Geophysics (Elsevier), pp. 305–353. https://doi.org/10.1016/S0065-2687(08)60176-4.

7. Forzieri, G., Alkama, R., Miralles, D.G., and Cescatti, A. (2017). Satellites reveal contrasting responses of regional climate to the widespread greening of Earth. Science *356*, 1180–1184. https://doi.org/10.1126/science.aal1727.

8. He, T., Liang, S., and Song, D. (2014). Analysis of global land surface albedo climatology and spatial-temporal variation during 1981–2010 from multiple satellite products. JGR Atmospheres *119*. https://doi.org/10.1002/2014JD021667.

9. Hollinger, D.Y., Ollinger, S.V., Richardson, A.D., Meyers, T.P., Dail, D.B., Martin, M.E., Scott, N.A., Arkebauer, T.J., Baldocchi, D.D., Clark, K.L., et al. (2010). Albedo estimates for land surface models and support for a new paradigm based on foliage nitrogen concentration. Global Change Biology *16*, 696–710. https://doi.org/10.1111/j.1365-2486.2009.02028.x.

10. D'Odorico, P., Bhattachan, A., Davis, K.F., Ravi, S., and Runyan, C.W. (2013). Global desertification: Drivers and feedbacks. Advances in Water Resources *51*, 326–344. https://doi.org/10.1016/j.advwatres.2012.01.013.

11. Mariano, D.A., Santos, C.A.C.D., Wardlow, B.D., Anderson, M.C., Schiltmeyer, A.V., Tadesse, T., and Svoboda, M.D. (2018). Use of remote sensing indicators to assess effects of drought and human-induced land degradation on ecosystem health in Northeastern Brazil. Remote Sensing of Environment *213*, 129–143. https://doi.org/10.1016/j.rse.2018.04.048.

12. Flanner, M.G., Shell, K.M., Barlage, M., Perovich, D.K., and Tschudi, M.A. (2011). Radiative forcing and albedo feedback from the Northern Hemisphere cryosphere between 1979 and 2008. Nature Geosci *4*, 151–155. https://doi.org/10.1038/ngeo1062.

13. Bright, R.M., Davin, E., O'Halloran, T., Pongratz, J., Zhao, K., and Cescatti, A. (2017). Local temperature response to land cover and management change driven by non-radiative processes. Nature Clim Change *7*, 296–302. https://doi.org/10.1038/nclimate3250.

14. Loranty, M.M., Berner, L.T., Goetz, S.J., Jin, Y., and Randerson, J.T. (2014). Vegetation controls on northern high latitude snow-albedo feedback: observations and CMIP 5 model simulations. Global Change Biology *20*, 594–606. https://doi.org/10.1111/gcb.12391.

15. Sturm, M., Racine, C., and Tape, K. (2001). Increasing shrub abundance in the Arctic. Nature *411*, 546–547. https://doi.org/10.1038/35079180.

16. Lobell, D.B., and Asner, G.P. (2002). Moisture Effects on Soil Reflectance. Soil Science Soc of Amer J *66*, 722–727. https://doi.org/10.2136/sssaj2002.7220.

17. Schaaf, C.B., Gao, F., Strahler, A.H., Lucht, W., Li, X., Tsang, T., Strugnell, N.C., Zhang, X., Jin, Y., Muller, J.-P., et al. (2002). First operational BRDF, albedo nadir reflectance products from MODIS. Remote Sensing of Environment *83*, 135–148. https://doi.org/10.1016/S0034-4257(02)00091-3.

18. Wang, Z., Schaaf, C.B., Chopping, M.J., Strahler, A.H., Wang, J., Román, M.O., Rocha, A.V., Woodcock, C.E., and Shuai, Y. (2012). Evaluation of Moderate-resolution Imaging Spectroradiometer (MODIS) snow albedo product (MCD43A) over tundra. Remote Sensing of Environment *117*, 264–280. https://doi.org/10.1016/j.rse.2011.10.002.

19. Liu, Y., Wang, Z., Sun, Q., Erb, A.M., Li, Z., Schaaf, C.B., Zhang, X., Román, M.O., Scott, R.L., Zhang, Q., et al. (2017). Evaluation of the VIIRS BRDF, Albedo and NBAR products suite and an assessment of continuity with the long term MODIS record. Remote Sensing of Environment *201*, 256–274. https://doi.org/10.1016/j.rse.2017.09.020.

20. Contribution of working group I to the sixth assessment report of the intergovernmental panel on climate change (2021). Climate change *2*, 2391.

21. Zhang, G., Xie, H., Fernandez, A., Kinnard, C., and Lhermitte, S. (2026). Remote sensing of the global cryosphere: Status, processes, and trends. Remote Sensing of Environment *334*, 115220. https://doi.org/10.1016/j.rse.2025.115220.

22. Hou, Z., Zhang, L., Peng, J., Forzieri, G., Jia, A., Xiao, Z., Qu, Y., Lin, J., Ji, D., Zhu, Z., et al. (2025). Radiative forcing reduced by early twenty-first century increase in land albedo. Nature *641*, 1162–1171. https://doi.org/10.1038/s41586-025-08987-z.

23. Schaaf, C.B., Gao, F., Strahler, A.H., Lucht, W., Li, X., Tsang, T., Strugnell, N.C., Zhang, X., Jin, Y., Muller, J.-P., et al. (2002). First operational BRDF, albedo nadir reflectance products from MODIS. Remote Sensing of Environment *83*, 135–148. https://doi.org/10.1016/S0034-4257(02)00091-3.

24. Hall, D.K., Riggs, G.A., and Salomonson, V.V. (1995). Development of methods for mapping global snow cover using moderate resolution imaging spectroradiometer data. Remote Sensing of Environment *54*, 127–140. https://doi.org/10.1016/0034-4257(95)00137-P.

25. Pang, G., Chen, D., Wang, X., and Lai, H.-W. (2022). Spatiotemporal variations of land surface albedo and associated influencing factors on the Tibetan Plateau. Science of The Total Environment *804*, 150100. https://doi.org/10.1016/j.scitotenv.2021.150100.

26. Muñoz-Sabater, J., Dutra, E., Agustí-Panareda, A., Albergel, C., Arduini, G., Balsamo, G., Boussetta, S., Choulga, M., Harrigan, S., Hersbach, H., et al. (2021). ERA5-Land: a state-of-the-art global reanalysis dataset for land applications. Earth Syst. Sci. Data *13*, 4349–4383. https://doi.org/10.5194/essd-13-4349-2021.

27. De Beurs, K.M., Wright, C.K., and Henebry, G.M. (2009). Dual scale trend analysis for evaluating climatic and anthropogenic effects on the vegetated land surface in Russia and Kazakhstan. Environ. Res. Lett. *4*, 045012. https://doi.org/10.1088/1748-9326/4/4/045012.

28. Fernandes, R., Brown, L., Canisius, F., Dash, J., He, L., Hong, G., Huang, L., Le, N.Q., MacDougall, C., Meier, C., et al. (2023). Validation of Simplified Level 2 Prototype Processor Sentinel-2 fraction of canopy cover, fraction of absorbed photosynthetically active radiation and leaf area index products over North American forests. Remote Sensing of Environment *293*, 113600. https://doi.org/10.1016/j.rse.2023.113600.

29. Ripple, W.J., Wolf, C., Mann, M.E., Rockström, J., Gregg, J.W., Xu, C., Wunderling, N., Perkins-Kirkpatrick, S.E., Schaeffer, R., Broadgate, W.J., et al. (2025). The 2025 state of the climate report: a planet on the brink. BioScience *75*, 1016–1027. https://doi.org/10.1093/biosci/biaf149.

30. Malmros, J.K., Mernild, S.H., Wilson, R., Tagesson, T., and Fensholt, R. (2018). Snow cover and snow albedo changes in the central Andes of Chile and Argentina from daily MODIS observations (2000–2016). Remote Sensing of Environment *209*, 240–252. https://doi.org/10.1016/j.rse.2018.02.072.

31. Yuan, S., Liu, Y., Qin, Y., and Zhang, K. (2023). Spatiotemporal variation of surface albedo and its influencing factors in northern Xinjiang, China. J. Arid Land *15*, 1315–1339. https://doi.org/10.1007/s40333-023-0069-5.

32. Seneviratne, S.I., Corti, T., Davin, E.L., Hirschi, M., Jaeger, E.B., Lehner, I., Orlowsky, B., and Teuling, A.J. (2010). Investigating soil moisture–climate interactions in a changing climate: A review. Earth-Science Reviews *99*, 125–161. https://doi.org/10.1016/j.earscirev.2010.02.004.

33. Ullah, W., Ahmad, K., Ullah, S., Tahir, A.A., Javed, M.F., Nazir, A., Abbasi, A.M., Aziz, M., and Mohamed, A. (2023). Analysis of the relationship among land surface temperature (LST), land use land cover (LULC), and normalized difference vegetation index (NDVI) with topographic elements in the lower Himalayan region. Heliyon *9*, e13322. https://doi.org/10.1016/j.heliyon.2023.e13322.

34. Hou, Z., Zhang, L., Peng, J., Forzieri, G., Jia, A., Xiao, Z., Qu, Y., Lin, J., Ji, D., Zhu, Z., et al. (2025). Radiative forcing reduced by early twenty-first century increase in land albedo. Nature *641*, 1162–1171. https://doi.org/10.1038/s41586-025-08987-z.

35. Yuan, S., Liu, Y., Liu, Y., Zhang, K., Li, Y., Enwer, R., Li, Y., and Hu, Q. (2024). Spatiotemporal variations of surface albedo in Central Asia and its influencing factors and confirmatory path analysis during the 21st century. International Journal of Applied Earth Observation and Geoinformation *134*, 104233. https://doi.org/10.1016/j.jag.2024.104233.